\documentclass[preprint,12pt]{elsarticle}
\usepackage{mathptmx}       
\usepackage{helvet}         
\usepackage{courier}        
\usepackage{type1cm}        
\usepackage{makeidx}         
\usepackage{graphicx}        
\usepackage{multicol}        
\usepackage[bottom]{footmisc}
\usepackage{amsmath}
\usepackage{amssymb}
\usepackage{bm}
\usepackage{subfigure}

\usepackage{amssymb}

\journal{Tacona}

\begin{document}

\begin{frontmatter}


 \title{Curved trajectories on transformed metal surfaces: Luneburg lens, beam-splitter, invisibility carpet and black hole for surface plasmon polaritons.}
 \author{Kadic Muamer\corref{cor1}}
\ead{muamer.kadic@fresnel.fr}
 \author{Dupont Guillaume}
 \author{Tieh-Ming Chang}
 \author{Sebastien Guenneau}
\author{Stefan Enoch}
 \cortext[cor1]{Corresponding author}
 \address{Institut Fresnel, CNRS, Aix-Marseille
Universit\'e,\\Campus universitaire de Saint-J\'er\^ome,
 13013 Marseille, France}


\begin{abstract}
Transformational optics are shown to markedly enhance the control of the electromagnetic wave trajectories
within metamaterials with unconventional functionalities such as a beam splitter, a toroidal carpet, a Luneburg
lens and a black hole, all of which are specially designed for surface plasmon polaritons propagating on a metal plate.
\end{abstract}


\end{frontmatter}


\section{Introduction} In 1998, a team led by Ebbesen discovered that resonant excitations of surface 
plasmons enhance electric fields at a surface that force light through its tiny holes, 
giving very high transmission coefficients in the sub-wavelength regime \cite{ebbesen}. 
Five years afterwards, Pendry, Martin-Moreno and Garcia-Vidal subsequently proposed a homogenized model 
of such structured metal surfaces in order to push the dispersion relation of surface plasmon polaritons 
to new boundaries \cite{science2004}. 
Plasmonics actually holds new promises with recent proposals in invisibility cloaks relying upon plasmonic metamaterials 
that have already led to fascinating results \cite{milton2,engheta,javier,baumeier}. 
These include resonant shells with a suitable out-of-phase polarizability in order to compensate the scattering 
from the knowledge of the electromagnetic parameters of the object to hide, and external cloaking, 
whereby a plasmonic resonance cancels the external field at the location of a set of electric dipoles. 
Recently, Baumeier et al. have demonstrated theoretically and experimentally that it is possible to reduce significantly 
the scattering of an object by a surface plasmon polariton, when it is surrounded by two concentric rings of 
point scatterers \cite{baumeier}. In this invited paper, we propose three plasmonic devices designed using powerful tools 
of transformational optics. We validate our theoretical proposals with three-dimensional computations using the finite element package COMSOL MULTIPHYSICS.
This emerging area of photonics is fueled by analogies with cosmologic physics 
in non-Euclidean space-time metrics \cite{hawking}: here, we focus our attention on a beam splitter, a toroidal carpet, a Luneburg lens 
and an optical black hole, for surface plasmons propagating at an anisotropic metal-dielectric interface. 
These four electromagnetic paradigms emphasize an unprecedented control of surface waves using mathematical 
tools of general relativity \cite{wheeler,philbin} in order to design some meta-surfaces \cite{spp1}-\cite{spp6} achieving
new plasmonic functionalities.

\section{Beam splitter}
\label{bs}
Recent advances in transformational optics have led to the proposal by de Rham et al. of a beam splitter which is a
heterogeneous anisotropic two-dimensional slab splitting an incident beam into two beams propagating along different directions.
Importantly, the slab is impedanced matched to the surrounding medium, hence it does not exhibit any
reflection at the entrance or exit boundaries \cite{rham}. We would like to adapt this device to the case of surface
plasmon polaritons propagating at a transformed metal-dielectric interface. The bottom-neck of the design is the
following transform \cite{rham}:   

\begin{equation}
\left\{
\begin{array}{c c l}
  x' & = & x\\
  y' & = & \dfrac{y_2 - y_1}{y_2} y + y_1 \\
  z' & = & z\\
\end{array}
\right. \quad \Rightarrow \quad
{\bf J}_{rr'} = \left( \begin{array}{c c c}
	    1 & 0 & 0 \\
	    c_{21} & \alpha^{-1} & 0 \\
	    0 & 0 & 1 \\
           \end{array} \right)
           \label{transfo1} 
\end{equation}
where ${\bf J}_{rr'}$ is the Jacobian matrix of the transformation, and $c_{12}$ is given by :

\begin{equation}
c_{21} = y_2 \dfrac{y - y_2}{(y_2 - y_1)^2} \dfrac{\partial y_1}{\partial x'} 
\end{equation}

\begin{figure}[h]
\begin{center}
\resizebox*{13cm}{!}{\includegraphics{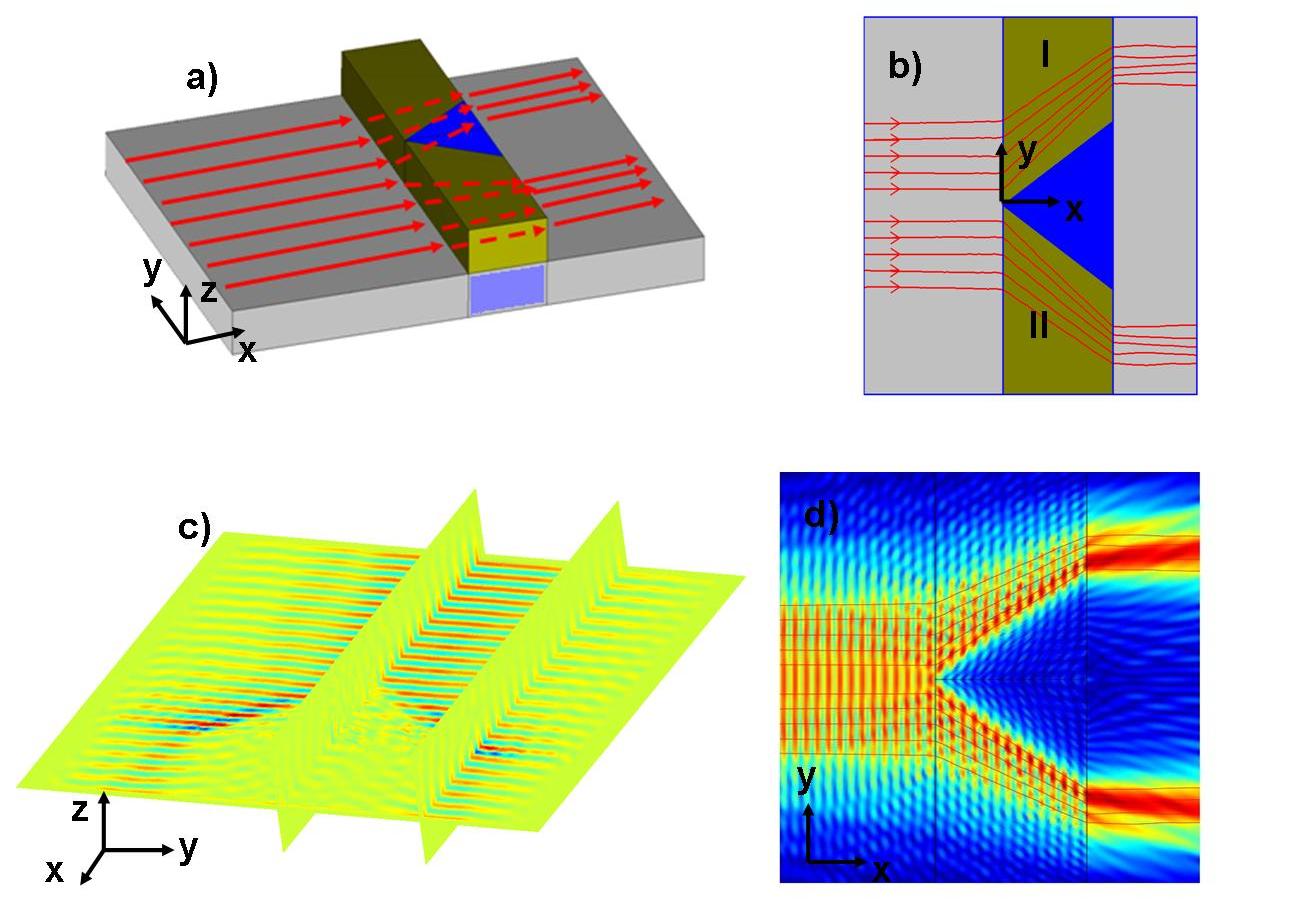}}
\\Figure 1: Beam splitter, a) Schematic diagram for a SPP beam splitter. A SPP is propagating from the left side.  b) Top view of beam splitter with
regions I and II deduced from the geometric transform \ref{transfo1} with $y_1=x$, $y_2=1$ in I and $y_1=-x$, $y_2=-1$ in II;
c) Three-dimensional plot of the real part of the magnetic field for a $y$-polarised SPP propagating in $x$-direction (phase); d) Two-dimensional plot of the normalised powerflow of the magnetic field in the $(x,y)$ plane. Stream lines appear in red color.
\end{center}
\label{fig1}
\end{figure}

The inverse transformation matrix takes the following form:

\begin{equation}
{\bf T}^{-1} = {\bf J}_{xx'}^{-1}{\bf J}_{xx'}^{-T} \hbox{det}({\bf J}_{xx'}) 
\end{equation}
where ${\bf J}_{xx'}^{-T}$ denotes the inverse transpose matrix of ${\bf J}_{xx'}$,
and $\hbox{det}({\bf J}_{xx'})$ its determinent. Besides, the explicit expression
of ${\bf T}^{-1}$ is:  
\begin{equation}
{\bf T}^{-1} = \left( \begin{array}{c c c}
	    \alpha^{-1} & -c_{21} & 0 \\
	    -c_{21} & \alpha (1 + c_{21}^2) & 0 \\
	    0 & 0 & \alpha^{-1}
          \end{array} \right) 
\end{equation}
where $\alpha = \dfrac{y_2 - y_1}{y_2}$. We show in figure 1 a typical computation
for a beam splitter in the particular case when the functions $y_1$ and $y_2$ bounding the regions
I and II are straight lines symmetric with respect to $y=0$ (here, $y_1=\pm x$ and $y_2=\pm 1$. We
however note that $y_1$ and $y_2$ could any smooth functions.

\section{Toroidal Carpet}
Next, we propose a design of a toroidal carpet which is deduced from the following transform:
\begin{equation}
\left\{
\begin{array}{c c l}
  x' & = & x\\
  y' & = & y \\
  z' & = & \dfrac{z_2 - z_1}{z_2} z + z_1\\
\end{array}
\right.
\end{equation}
where $\mathbf{J}_{zz'}$ is the Jacobian matrix of the transformation as per:
\begin{equation}
{\bf J}_{zz'} = \left( \begin{array}{c c c}
            1 & 0 & 0 \\
	    0 & 1 & 0 \\
	    \dfrac{\partial z}{\partial x'} & \dfrac{\partial z}{\partial y'} & \alpha^{-1} \\
           \end{array} \right)
\end{equation}
and partial derivatives can be expressed as:\\

\noindent $\dfrac{\partial z}{\partial x'} = z_2 \dfrac{z' - z_2}{(z_2 - z_1)^2} \dfrac{\partial z_1}{\partial x'} +
	z_1 \dfrac{z_1 - z'}{(z_2 - z_1)^2} \dfrac{\partial z_2}{\partial x'} \quad ; \quad$
$\dfrac{\partial z}{\partial y'} = z_2 \dfrac{z' - z_2}{(z_2 - z_1)^2} \dfrac{\partial z_1}{\partial y'} +
	z_1 \dfrac{z_1 - z'}{(z_2 - z_1)^2} \dfrac{\partial z_2}{\partial y'}$
	
The resulting transformation matrix is:
\begin{equation}
{\bf T}^{-1} =  \left( \begin{array}{c c c}
            \alpha^{-1} & 0 & -\dfrac{\partial z}{\partial x'} \\
	    0 & \alpha^{-1} & -\dfrac{\partial z}{\partial y'} \\
	    -\dfrac{\partial z}{\partial x'} & -\dfrac{\partial z}{\partial y'} &
		  \alpha \left( 1+ \left( \dfrac{\partial z}{\partial x'} \right)^2 +\left( \dfrac{\partial z}{\partial y'} \right)^2 \right)\\
           \end{array} \right)
\end{equation}
	
\bigskip

In the design shown in figure 2, we actually mapped a toroidal bump (described by the surface of altitude $z_1$) on a flat ring (described by the surface $z_2$):\\
  
$z_1 = \sqrt{b^2-(\sqrt{x^2+y^2}-a)^2}+zo$ $\quad$ with $\quad$ $a = 0.3 \;;\; b = 0.1 \;;\; zo = -0.05$\\
\bigskip
$z_2 = 0.2$\\
\bigskip
$\dfrac{\partial z_1}{\partial x} = -\dfrac{(\sqrt{x^2+y^2}-a)x}{\sqrt{b^2-(\sqrt{x^2+y^2}-a)^2}.\sqrt{x^2+y^2}}$ $\quad ; \quad$
$\dfrac{\partial z_1}{\partial y} = -\dfrac{(\sqrt{x^2+y^2}-a)y}{\sqrt{b^2-(\sqrt{x^2+y^2}-a)^2}.\sqrt{x^2+y^2}}$
\bigskip

We show in figure 2 the result of our three-dimensional computations, which clearly demonstrate that an electromagnetic field can be stored within an invisible carpet
that does not perturb the ambient SPP field.
Applications might be in hard-discs for an all-optic computer.
\begin{figure}[h]
\begin{center}
\resizebox*{13cm}{!}{\includegraphics{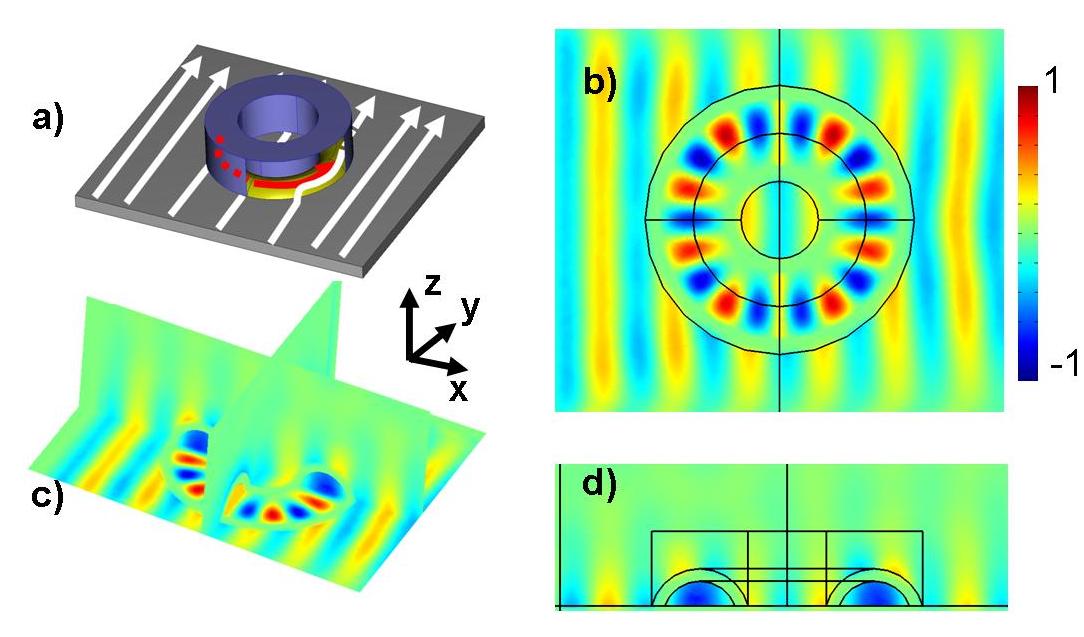}}
\\Figure 2: Plane wave, a) Electromagnetic field incident from the top ($z$-axis), with the magnitude of the magnetic field being represented; b) Phase representation of the magnetic field.
$y$-polarized SPP wave; c) Two-dimensional plot in the $(x,y)$ plane; d) Two-dimensional plot in the $(x,y)$ plane;
Moreover, an electromagnetic field has been launched inside the toroidal carpet filled here with a dielectric medium of permittivitty $\varepsilon=2$ and surrounded by a thin metal coating.
The field circulates in closed trajectories and does not perturb the SPP propagation on the metal plate.
\label{fig2}
\end{center}
\end{figure}

\section{Luneburg lens}
There is currently a renewed interest in gradient index lenses, such as Eaton and Luneburg lenses and Maxwell's fisheye. Following the earlier proposal
by Liu et al. \cite{spp3} of a Luneburg lens which is made of a homogeneous dielectric with a specific spatial variation of the profile, we decide here
to investigate the case of a heterogeneous Luneburg lens, with the usual spatially varying refractive index given by \cite{luneburg}: 
\begin{equation}
n=\sqrt{2-(\sqrt{x^2+y^2}/R)^2} \; ,
\end{equation}
where $R$ is the radius of the lens and the coordinate axis is located at the center of the lens.
\begin{figure}[h]
\begin{center}
\resizebox*{7cm}{!}{\includegraphics{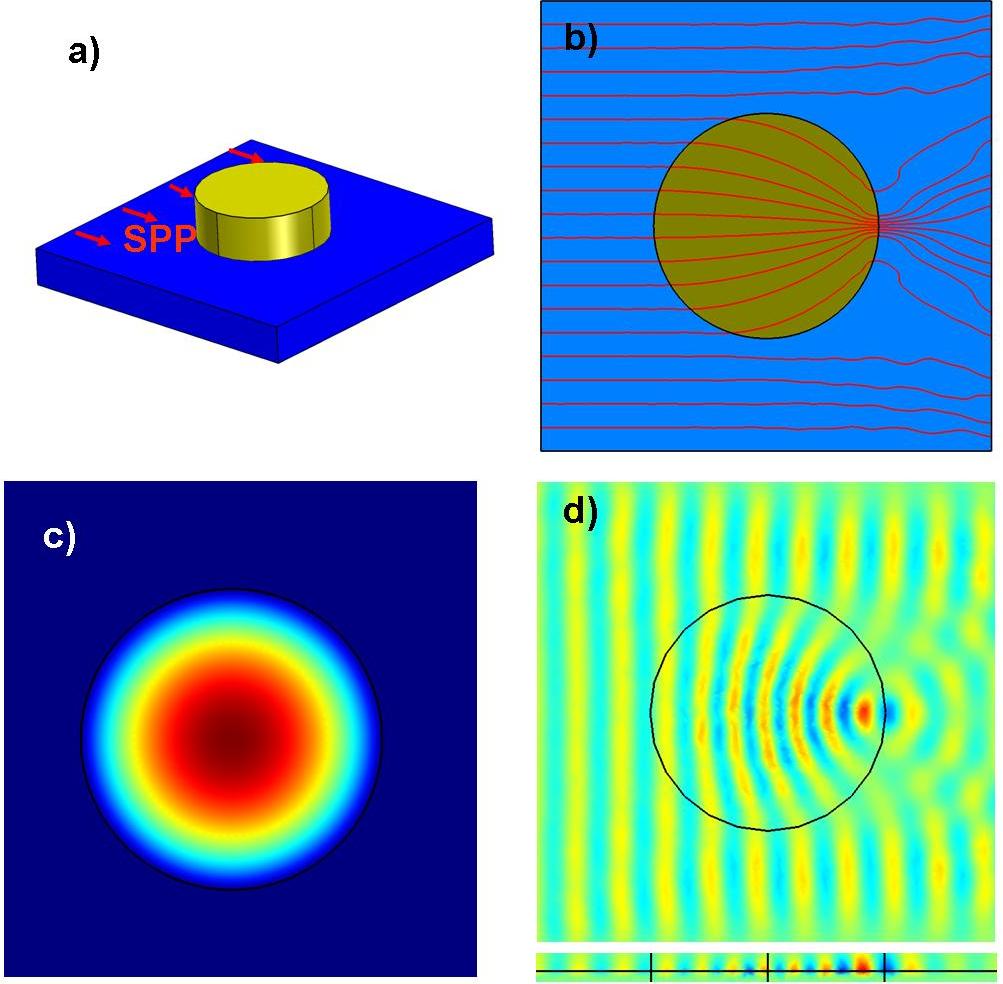}}
\\ Figure 3: Plasmonic cylindrical Luneburg lens with a spatially varying refractive index.
(a) Schematic diagram for a cylindrical Luneburg lens on a metal plate (the color scale ranges from $1$ (vacuum), blue color, to $1.414$, red color; (b) Streamlines show the focussing effect for a SPP incident upon the Luneburg lens
(view from above); (c) Spatial distribution of the refractive index;  (d) Top view and side view for a $y$-polarized SPP propagating from the left and incident upon the Luneberg
lens at $700$ nm (2D plots of the normalized real part of the magnetic field in the $(x,y)$ and $(x,z)$ planes). 
\label{fig3}
\end{center}
\end{figure}

We note that the metal is kept untounched, and no magnetism is involved in this design which thus looks as a reasonable design for an experimental vailidation.

\section{Black hole}

\begin{figure}[h]
\begin{center}
\resizebox*{12cm}{!}{\includegraphics{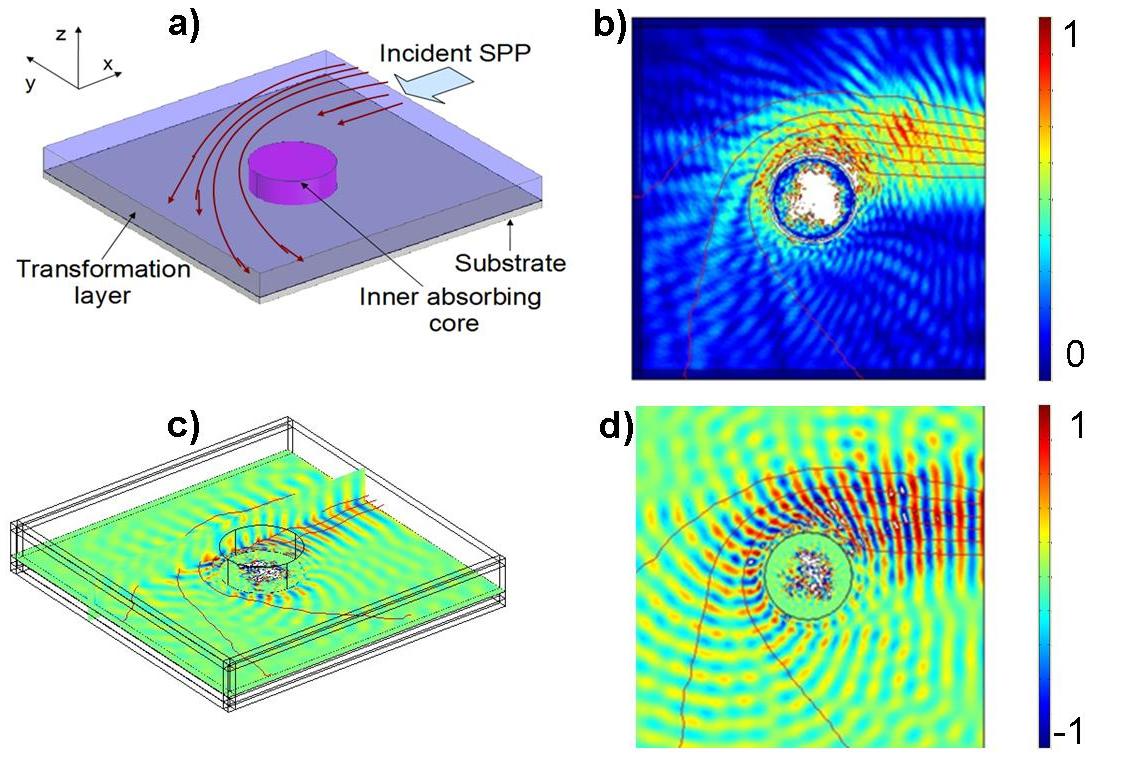}}
\\Figure 4: Plasmonic cylindrical black hole mimicking a Schwarschild metric. (a) Schematic diagram for a transformed metal plate
with a cylindrical transformed medium on the top; (b,c,d) A SPP Gaussian beam making an angle of $60$ degrees with the $x$-axis is incident
from the right (b: top view, normalized magnitude of magnetic field; c: 3D plot for the normalized real part of magnetic field; d:
top view phase of magnetic field).
\label{fig4}
\end{center}
\end{figure}

The infinitesimal line element in a Schwarschild metric for a black hole can be expressed as \cite{wheeler}
\begin{equation}
ds^2=\left( 1 - \frac{L}{r})c^2dt^2\right) - {\left( 1 - \frac{L}{r} \right)}^{-1} dr^2 - r^2 d\theta^2 - dz^2 \; , 
\end{equation}
where $L$ denotes the event horizon \cite{hawking}, $t$ is the time variable, $(r,\theta,z)$ denote the space coordinates
and $c$ is the velocity of light. Without loss of generality, one can simply skip the time variable within the framework
of time-harmonic Maxwell's equations.

Following the proposal of an optical black hole by Chen et al. \cite{chen}, we rederived the transformation matrix associated
with the spatial part of the Schwarschild metric and we found that
\begin{equation}
{\bf T}^{-1}= \dfrac{A}{1-\dfrac{L_1}{r}} \left( \begin{array}{c c c}
            1-\dfrac{{L_x}^2}{r^3} & -\dfrac{L_{xy}}{r^3} & 0 \\
	   -\dfrac{L_{xy}}{r^3}  & 1-\dfrac{{L_y}^2}{r^3} & 0 \\
	    0 & 0 & 1\\
           \end{array} \right)        
\end{equation}

\noindent where $A = 1-i$ in the absorbing core ($r \leq L$) and $A = 1$ in the transformation layer ($r>L$).

In the present case, we launch a SPP at the wavelength $\lambda = 700 \, nm$ and width $1400 \, nm$. The 
radius of the inner absorbing core is $L = 0.9 \, \mu m$ and $L_1 = 1 \, \mu m$. In figure 4, the trajectory of the Gaussian
beam is clearly bent around the center part of the black hole (see also the streamlines), as it should in a Schwarschild metric. 
           

\section{Conclusion} In this paper, we reviewed some techniques of transformational optics applied to surface plasmon polaritons propagating at a flat interface separating a transformed Drude metal and a transformed dielectric medium. We have proposed four original meta-surfaces: A beam splitter which splits a large Gaussian SPP in two narrower Gaussian SPPs with different directions; A toroidal carpet which detours an SPP without disturbing its wavefront and its amplitude; A Luneburg lens which focusses an incident SPP onto a small region behind the lens; An optical black hole which can be modeled as vacuum solutions to the Einstein field equations. The latter device can be seen as intrinsic parts of the unbounded version of the Schwarzschild metric describing an eternal black hole with no charge and no rotation. Indeed, all geodesics of a free-falling particle in the spacetime can be continued arbitrarily far into the particle's future or past, unless the geodesic hits a gravitational singularity like the one at the center of the black hole's interior. In a similar way, we have seen that ray trajectories of a time-harmonic solution to the Maxwell's equations in the optical black hole converge towards the center of the black hole, in some kind of spiral manner, provided the rays intersect the event horizon of the black hole.

We discovered while finalising the mansucript that a paper had been submitted by Zhang's team on numerical and experimental results for a plasmonic Luneberg lens, but we emphasize that our results cross-check their simulations.


\begin{thebibliography}{23}
\bibitem{ebbesen} T. W. Ebbesen, H. J. Lezec, H. F. Ghaemi, T. Thio, P. A. Woff, Nature 391, 667 (1998). 
\bibitem{science2004} J.B. Pendry, L. Martin-Moreno and F.J. Garcia-Vidal, Mimicking surface plasmons with structured surfaces, Science {\bf 305}, 847 (2004) 
\bibitem{milton2} N.A. Nicorovici, R.C. McPhedran and G.W. Milton,``Optical and dielectric properties of partially resonant composites,'' Phys. Rev. B {\bf 49}, 8479-8482 (1994). 
\bibitem{engheta} A. Alu and N. Engheta, ``Achieving transparency with plasmonic and metamaterial coatings,'' Phys. Rev. E {\bf 72} 016623 (2005). 
\bibitem{javier} F.J. Garcia de Abajo, G. Gomez-Santos, L.A. Blanco, A.G. Borisov and S.V. Shabanov, Physical Review Letters {\bf 95} 067403 (2005). 
\bibitem{baumeier} B. Baumeier, T.A. Leskova and A.A. Maradudin, ``Cloaking from surface plasmon polaritons by a circular array of point scatterers,'' Physical Review Letters {\bf 103}, 246809 (2009) 
\bibitem{hawking} S. Hawking and G. Ellis, The Large Scale Structure of Space-Time, Cambridge University Press (1973) 
\bibitem{wheeler} J.A. Wheeler, Geometrodynamics. New York: Academic Press (1962)
\bibitem{philbin}
U. Leonhardt and T.G Philbin, ``General relativity in electrical engineering,'' New J. Phys. 8 247 (2006)
\bibitem{spp1}
I.I. Smolyaninov, ``Transformational optics of plasmonic metamaterials,'' New J. Phys. 10(11), 115033 (2008).
\bibitem{spp2}
P.A. Huidobro, M. L. Nesterov, L. Martin-Moreno, and F. J. García-Vidal,``Transformation Optics for
Plasmonics,'' Nano Lett. 19, 1985 (2010).
\bibitem{spp3}
Y. Liu, T. Zentgraf, G. Bartal, and X. Zhang, ``Transformational Plasmon Optics,''
Nano Lett. 6, 1991–1997 (2010).
\bibitem{spp4}
I.I. Smolyaninov, V. N. Smolyaninova, A. V. Kildishev, and V. M. Shalaev,``Anisotropic metamaterials
emulated by tapered waveguides: application to optical cloaking,'' Phys. Rev. Lett. 102(21), 213901 (2009).
\bibitem{spp5}
M. Kadic, S. Guenneau, and S. Enoch, ``Transformational plasmonics: cloak, concentrator and rotator for SPPs,''
Opt. Express 18(11), 12027-12032 (2010).
\bibitem{spp6}
J. Renger, M. Kadic, G. Dupont, S.S. Acimovic, S. Guenneau, R. Quidant, and S. Enoch,
``Hidden progress: broadband plasmonic invisibility,'' Optics Express 18(15), 15757-15768 (2010) 
\bibitem{rham}
M. Rahm, D.A. Roberts, J.B. Pendry, and D.R. Smith,
``Transformation-optical design of adaptive beam bends and beam expanders,''
Optics Express 16(15), 11555-11567 (2008)        
\bibitem{zheludev}
M. Bashevoy, V. Fedotov, and N. Zheludev, ``Optical whirlpool on an absorbing metallic nanoparticle,'' Opt. Express 13, 8372-8379 (2005)
\bibitem{luneburg} 
R.K. Luneburg, ``Mathematical Theory of Optics'', Providence, Rhode Island: Brown University, 189–213 (1944).
\bibitem{chen}
H. Chen, R-X Miao and M. Li,
Transformation optics that mimics the system oustide a Schwarzschild black hole,
Opt. Express 15183-15188 (2010).
\bibitem{spp7}
T. Zentgraf, Y. Liu, M.H. Mikkelsen, J. Valentine and X. Zhang,
Plasmonic Luneburg and Eaton lenses,
Nature Nanotechnology (DOI: 10.1038/NNANO.2010.282)
\end{thebibliography}







\end{document}